\documentclass[
 reprint,
 superscriptaddress,
 amsmath,
 amssymb,
 aps,
]{revtex4-2}

\usepackage{graphicx}
\usepackage{dcolumn}
\usepackage{bm}
\usepackage[caption=false]{subfig}
\usepackage{sidecap}
\usepackage{float}
\usepackage{physics}
\usepackage{siunitx}

\newcommand{\ch}{{\text{Ch}_{1}}}

\begin{document}

\preprint{APS/123-QED}

\title{Topological phases in interacting spin-1 systems}

\author{A. Alnor}
\affiliation{
Department of Physics and  Astronomy, Aarhus University, DK-8000 Aarhus C,  Denmark
}
\author{T. B{\ae}kkegaard}
\affiliation{
Department of Physics and  Astronomy, Aarhus University, DK-8000 Aarhus C,  Denmark
}
\affiliation{
Institute of Photonics, Leibniz University Hannover, Nienburgerstr. 17, 30519 Hannover
}

\author{N. T. Zinner}
\affiliation{
Department of Physics and  Astronomy, Aarhus University, DK-8000 Aarhus C,  Denmark
}
\affiliation{
Aarhus Institute of Advanced Studies, Aarhus University, DK-8000 Aarhus C,  Denmark
}

\date{\today}

\begin{abstract}
Different topological phases of quantum systems has become areas of increased focus in recent decades. In particular, 
the question of how to realize and manipulate systems with non-trivial first Chern number is pursued both experimentally 
and theoretically. Here we go beyond typical spin 1/2 systems and consider both single and coupled spin 1 systems as 
a means of realizing higher first Chern numbers and studying the emergence of different topological phases. 
We show that rich topological phase diagrams can be realized by coupling two spin 1 systems using both 
numerical and analytical methods. Furthermore, we consider a concrete realization of spin 1 systems using a 
superconducting circuit. This realization includes non-standard spin-spin interaction terms that may endanger
the topological properties. We argue however that the realistic circuit Hamiltonian including all terms should 
be expected to show the rich phase structure as well. This puts our theoretical predictions within reach of state-of-the-art experimental setups. 
\end{abstract}

\maketitle

\section{Introduction}
\label{chap:intro}

The phenomena of topological properties of quantum systems have seen much research in recent years as a tool for describing global non-trivial features.
Derived in 1950 by T. Kato \cite{Kato}, though named after M.V. Berry who discovered it independently in 1984 \cite{berry84}, the Berry phase, also called the geometric phase, arises by adiabatically varying the parameters governing the quantum system.
This gives rise to the Berry curvature, whose closed 2-D surface integral is an integer multiple of $2\pi$. This integer is the first Chern number of the system, a topological invariant \cite{thouless1982quantized,kohmoto1985topological,wen1995topological,schnyder2008classification,kitaev2009periodic}.
This connection between varying external parameters and global features of the system, gives a method of experimentally probing the topological features of quantum systems \cite{hasan2010colloquium,qi2011topological}.
Two regions in the space of external parameters with differing ground-state topological properties, i.e. differing topological quantum numbers, are considered as distinct topological phases \cite{Chiu16}. These topological phases are considered as quite stable since the transitions between the two can only happen by closing the energy gap between the ground state and the first excited state. These distinct topological phases, and the dependence on externally changing parameters will be the focus of our attention.

Topological properties of quantum systems are of increasing interest in condensed matter physics, due to exhibiting interesting properties that are typically not seen in classical systems. Some of these properties are believed to have technological applications, including their use in quantum information technology, topological insulators and unconventional superconductors \cite{nayak2008non,pachos2012introduction,bernevig2013topological}. 
For instance, for two-dimensional filled bands, and three-dimensional Fermi surfaces, the Hall conductance of the system is characterized by the Chern number \cite{bernevig2013topological}.

Due to the difficulty of probing topological properties in physical systems directly, emulation of the systems by other means is sometimes employed. In 1982, Richard Feynman proposed to use quantum simulation to investigate otherwise inaccessible quantum phenomena \cite{Feynman1982}. In recent years, there have been extended efforts to perform quantum simulation \cite{georgescu2014quantum} of systems that 
exhibit non-trival Chern numbers using several different platforms including cold atomic gases \cite{zhang2018topological,RevModPhys.91.015005}, Rydberg systems \cite{adams2019rydberg,browaeys2020many}, trapped ions \cite{blatt2012quantum,RevModPhys.93.025001}, and photonics \cite{lu2016topological,ozawa2019topological}.
Here we will focus on the platform of superconducting circuits \cite{xiang2013hybrid,gu2017microwave,krantz2019quantum,kjaergaard2020superconducting,blais2021circuit} as our main platform for studying systems that may realize non-trivial first Chern numbers. 
Inspired by two papers from 2014 \cite{schroer, roushan}, in which the Chern number was experimentally measured in artificial spin-$1/2$ qubit systems of quantum circuits, we will consider here how higher spins can realize topological phases. In particular, 
we will study spin 1 systems, also known as qutrits. These have been the focus of attention for more than a decade within the realm of superconducting circuits \cite{Neeley722,dicarlo2009demonstration,bianchetti2010control,abdumalikov2013experimental,peterer2015coherence,Tan18,baekkegaard2019realization,loft2020quantum,zhang2021synthesizing}. 

Here we explore how to obtain non-trivial Chern numbers in spin systems realized as a quantum simulation in the sense that they do not represent real particles, but rather achieve the same dynamics using superconducting circuits. We will explore topological phases in three types of spin-$1$ systems. First a simple spin-$1$ system controlled by a magnetic field. Second a system with two coupled qutrit subsystems controlled by a single coupling parameter. Finally, we will look at an example of how this could be implemented in a system of qutrits using superconducting circuits that was suggested in a recent paper \cite{rasmussen2021superconducting, Baekkegaard} using three different coupling parameters. The analysis will be based on both analytic solutions and numerical simulation using the QuTiP simulation package \cite{johansson2012qutip}. Comments are placed throughout on the aspects of superconducting circuits that might be relevant for physical realization. The exposition below will start by consider the systems as abstract theoretical models and then we proceed to discuss the concrete setups within superconducting circuits.

\section{Formalism}
\label{chap:berry}

In this section we introduce the basic theoretical framework and explain our notation. We will make use of spin operators. These will be denoted $S^i$ for $i=x,y,z,+,-$. The spin vector is defined as $\textbf{S}=(S^x,S^y,S^z)$. In case of an operator $S$ working on a particular system $j$, we will use the notation $S_j$. As an example, the $z$-component of the spin for the first particle is $S_1^z$. When dealing with spin-1 three-level systems, the states will be given in the basis $\left\{ \uparrow,0,\downarrow \right\}$ and the systems will be referred to as qutrits.

\subsection{Brief introduction to Berry's phase and related concepts}
The derivation of Berry's phase,  and of the related Chern number, can be found many places in literature, see for example chapter 2 of \cite{bernevig2013topological} or chapter 5 of \cite{sakurai}. We will simply state the main results. 

Adiabatic time evolution is defined as time evolution where the parameters of the system are changed slowly, allowing the system to adapt to the change. The adiabatic theorem loosely states, that in the adiabatic limit, a system initially being in the $n$th eigenstate, will be in the $n$th eigenstate of the new system. 
If one considers situations where the parameters of the Hamiltonian will be changed slowly and return to the initial ones, one might expect that the system will then be identical to the original system, however, a phase may be picked up.

If a system starts out in the eigenstate $|n\rangle$ of $H(0)$, then it remains in the eigenstate $|n\rangle$ of $H(t)$, and the total state can be written
\begin{equation}
\ket{\Psi(t)}=e^{i \gamma_{n}(t)} e^{i \theta_{n}(t)}\ket{n(t)}.
\end{equation}
where
\begin{equation}
    \gamma_n(t) = \int_0^t \left\langle n(t') \left|\frac{\partial}{\partial t'} \right| n(t') \right\rangle \dd{t'}
\end{equation}
is Berry's phase and $\theta_n$ is the dynamic phase, which is not of interest here.

\subsection{Degeneracies and Chern numbers}
We let $\mathbf{R}$ be a vector in the parameter space, i.e. the components that control the Hamiltonian are collected in this vector. 
The Berry phase gives rise to the Berry curvature $\mathbf{F}_{n}$, a local (in $\mathbf{R}$-space) property of the quantum system, which in the three-dimensional case can be calculated as
\begin{equation} \label{eq:CurvNab}
\mathbf{F}_{n}(\mathbf{R})= -\text{Im} \sum_{m \neq n} \frac{\left\langle n\left|\nabla_{\mathbf{R}} H \right| m\right\rangle \times\left\langle m\left|\nabla_{\mathbf{R}} H\right| n\right\rangle}{\left(E_{m}-E_{n}\right)^{2}}.
\end{equation}
As can be seen from eq. \eqref{eq:CurvNab}, degeneracies are the source of Berry curvature due to the divergence that will result from them. 

The integral of the Berry curvature over a closed surface is quantized in integer multiples of $2\pi$ \cite{BSimon}, the integer is the first Chern number of the system, 
\begin{equation}
   \text{Ch}_1 = \frac{1}{2\pi}  \oint_{\mathcal{S}} \mathbf{F}\cdot \dd\mathbf{S}.
\end{equation}
where $\mathbf{S}$ is the normal vector of the surface $\mathcal{S}$ spanned by the closed loop in parameter space. The first Chern number is often simply called \textit{the} Chern number and will be referred to as such throughout this text. The Chern number is a topological invariant. In this case, it reveals the number of enclosed degeneracies. 

The Berry curvature is sometimes described as an effective magnetic field with degeneracies acting as the source of the field, equivalent to magnetic monopoles (see for instance \cite{Jackiw}). In this framework, the Chern number is simply found using Gauss' law \cite{GriffED} and the number is signalling the enclosed degeneracies with their monopole charge.
In a similar vein, Gritsev et al. showed that the Berry curvature produces an effective force \cite{Gritsev6457} in a system initially in the ground state driven by a slow linear parameter change. By changing the parameters in the $\mu$-direction with rate $v_{\mu} = \partial \mu/\partial t$, the system, with time dependent wave function $\ket{\Psi(t)}$, experiences a generalized force in the $\nu$-direction $M_\nu=-\partial H/\partial \nu$, related to the $\mu\nu$ component of the Berry curvature
\begin{equation} \label{eq:forcecurve}
    \expval{M_\nu(t)} \equiv \expval{M_\nu}{\Psi(t)} = \expval{M_\nu}{0} + v_{\mu} F_{0, \mu \nu} + \mathcal{O}\left(v_{\mu}^2\right)
\end{equation}
where $\mathcal{O}\left(v_\mu^2\right)$ are higher order terms that will be neglected, due to the change happening slowly. The first term is the value in the adiabatic limit $v_\mu = 0$ in which case the system stays in the exact ground state at all times. Thus, we can directly measure the Berry curvature from the linear response by measuring $\expval{M_\nu}$. Both prescriptions of the curvature are used throughout this paper, eq. (\ref{eq:CurvNab}) when making analytical predictions, eq. (\ref{eq:forcecurve}) when performing simulations.

\section{Three-level systems}
\label{chap:three}

As a warm-up to the coupled systems and an introduction to how we treat the system analytically, we start by considering a relatively generic spin 1, or three-level system. We follow previous discussion of similar systems with spin 1/2 \cite{schroer,roushan}, as well as the study of \cite{Tan18} where topological band structure has been investigated with a superconducting circuit and achieved the measurement of a Chern number greater than unity.

The spin matrices for spin 1 will be taken to be of the form
\begin{widetext}
\begin{equation}
    S^x = \frac{\hbar}{\sqrt{2}}
    \begin{pmatrix}
    0 & 1 & 0 \\
    1 & 0 & 1 \\
    0 & 1 & 0
    \end{pmatrix}, \, 
    S^y = \frac{i \hbar}{\sqrt{2}}
    \begin{pmatrix}
    0 & -1 & 0 \\
    1 & 0 & -1 \\
    0 & 1 & 0
    \end{pmatrix}, \,
    S^z = \hbar 
    \begin{pmatrix}
    1 & 0 & 0\\
    0 & 0 & 0\\
    0 & 0 & -1
    \end{pmatrix} \text{ and } S^+ = {\hbar \sqrt{2}}
    \begin{pmatrix}
    0 & 1 & 0\\
    0 & 0 & 1\\
    0 & 0 & 0
    \end{pmatrix} 
    = \left(S^-\right)^\dagger
\end{equation}
\end{widetext}
From this point on, where units in which $\hbar=1$ are used.

\subsection{Simple three-level system}\label{sec:simple}
We first take a purely analytical approach to the simple three-level system. This procedure is identical to the one performed by Berry in \cite{berry84}.
The Hamiltonian we will be dealing is that of a spin 1 in a magnetic field, and has the form 
\begin{equation}
	H = \textbf{R}\cdot \textbf{S},
\end{equation}
where $\textbf{R}$ is the vector in parameter space. The Berry curvature is calculated using eq. (\ref{eq:CurvNab})
Explicitly calculating its eigenvalues, we see that it has energies $E_\uparrow=-E_\downarrow = |\textbf{R}| = R$ and $E_0=0$. 
A quick calculation gives $\left(E_{\uparrow/\downarrow}-E_{\downarrow/\uparrow}\right)^2= 4 R^2$ and $\left(E_{\uparrow/\downarrow}-E_0\right)^2 = \left(E_0-E_{\uparrow/\downarrow}\right)^2 = R^2$, while $\nabla_\textbf{R} H= \textbf{S}$.
Since the Hamiltonian is spherically symmetric, we can fix $\textbf{R}$ to point in the $\hat{\textbf{z}}$ direction. We can then write $\textbf{S}$ as
\begin{equation}
	\textbf{S}=\frac{1}{2}\left(S^+ + S^-\right)\hat{\textbf{x}} + \frac{1}{2i}\left(S^+ - S^-\right)\hat{\textbf{y}} +S_z \hat{\textbf{z}}.
\end{equation}
Explicit calculation of the Berry curvature by eq. (\ref{eq:CurvNab}) yields, by invoking the requirement for spin eigenvalues, and applying orthogonality
\begin{equation}
	\textbf{F}_\uparrow 
	= - \frac{\hat{\textbf{z}}}{R^2} .
\end{equation}
Remembering that we had rotated our axes to point in the $\hat{\textbf{z}}$ direction, we arrive at the expression
\begin{equation}
\textbf{F}_\uparrow = - \frac{\hat{\textbf{R}}}{R^2},
\end{equation}
while similar calculation will result in
\begin{equation} \label{eq:chnul}
\textbf{F}_\downarrow = \frac{\hat{\textbf{R}}}{R^2}, \quad \textbf{F}_0 = \textbf{0}.
\end{equation}
The integral of $\textbf{F}_{\uparrow/\downarrow}$ is well-known in electrodynamics \cite{GriffED}, and the integral over any surface enclosing $\textbf{0}$ will yield
\begin{equation} \label{eq:fipi}
	\oint_\mathcal{S} \textbf{F}_\uparrow \cdot d\textbf{a} = -4 \pi.
\end{equation} 
So that, up to a sign, we arrive at $\ch = 2$, when the degeneracy is enclosed. This degeneracy is known as the Weyl point and in this case has topological charge $2$.

\subsubsection{Simulating the simple system}
\begin{figure}
    \centering
    \includegraphics[width=0.8\columnwidth]{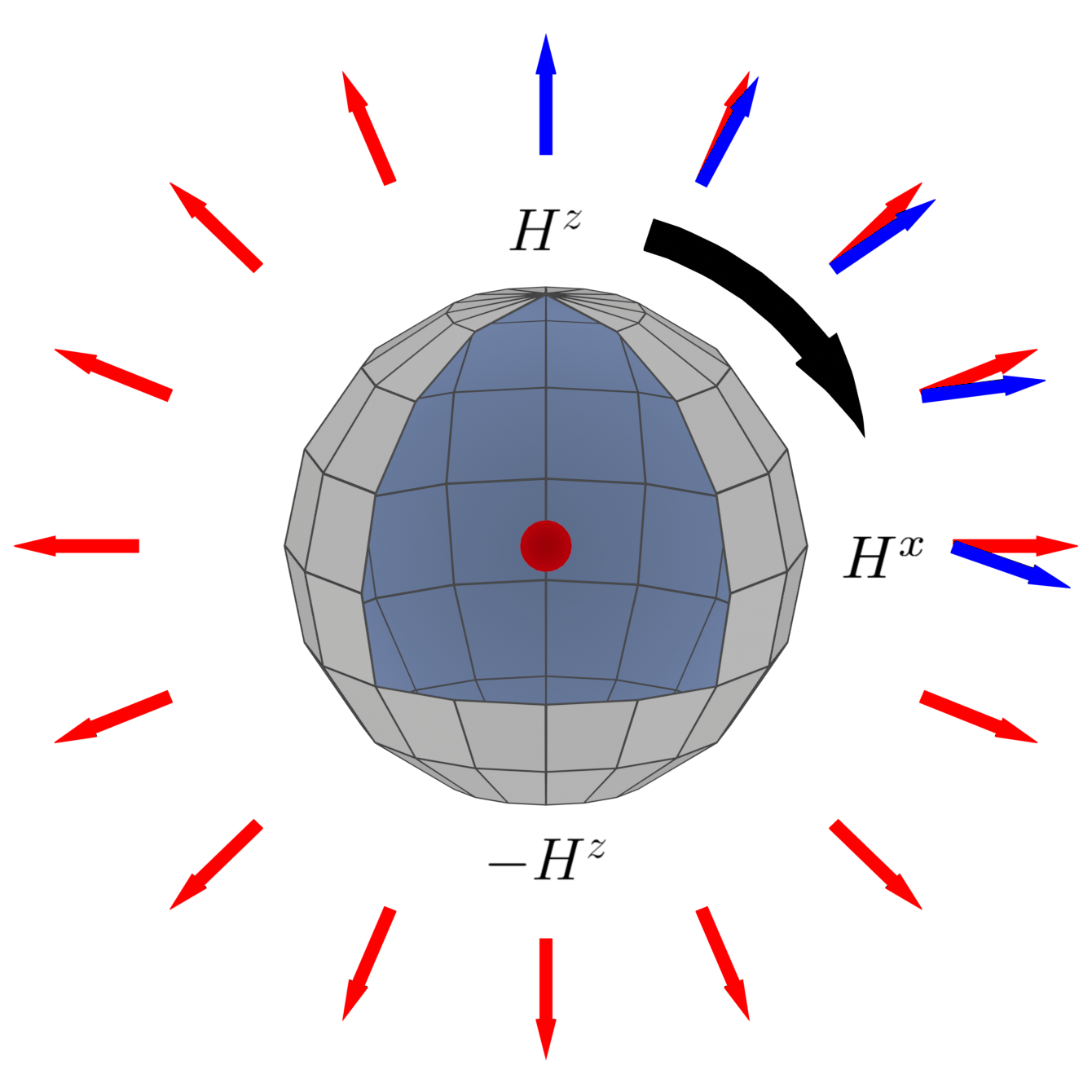}
    \caption{Schematic of the path traversed in the parameter space over the closed spherical manifold described by $\textbf{H}_\textbf{r}$. Starting at $\textbf{H}_\textbf{r}(t=0)=H^z\hat{\textbf{z}}$ and moving in a semicircle to $\textbf{H}_\textbf{r}(t=t_\text{ramp}) = -H^z\hat{\textbf{z}}$.
    The adiabatic case being represented by red arrows and the measured deflection, caused by the ramping, being represented by blue arrows. Inside the sphere, lies the Weyl point causing the deflection from adiabaticity which is used to extract the Chern number. This figure is based on fig. 1 in \cite{roushan}.}
    \label{fig:path}
\end{figure}

In order to address the potential conditions of an experiment with a 
three-level system, we now simulate a protocol for measuring the Chern number by use of eq. \eqref{eq:forcecurve}. We consider Hamiltonians of the form
\begin{equation}
    H= -\left(H_{0} S^z + \textbf{H}_\textbf{r}\cdot\textbf{S} \right)
\end{equation}
with $\textbf{H}_\textbf{r}=(H^x,H^y,H^z)=H_\textbf{r}\left( \sin(\theta)\cos(\phi), \sin(\theta)\sin(\phi), \cos(\theta) \right)$. For all simulations we use Hamiltonian parameters 
that are within the usual set of values for superconducting circuits \cite{rasmussen2019controllable,barfknecht2019realizing,rasmussen2020simple}, and we set $H_\textbf{r}=10\cdot 2\pi$ MHz.
The system is initialized in its instantaneous ground state for a particular value of $H_0$. We then change (ramp) the value of $\theta$ from $0$ to $\pi$ as $\theta(t)=\pi t/t_\text{ramp}$, finally stopping at $t=t_\text{ramp}$. We denote
\begin{equation}
    v_\theta = \frac{\partial \theta}{\partial t}
\end{equation}
By ramping $\theta$ like so, we create a curve through parameter space in the shape of a semicircle from $H_\textbf{r}S^z$ to $-H_\textbf{r}S^z$ (see fig. \ref{fig:path}).

Our Hamiltonian is cylindrically symmetric around the $z$-axis and as such, the Berry curvarture is independent of $\phi$. We therefore fix $\phi(t)=0$ throughout. 
The expectation value of the generalized force of the system is then
\begin{equation}
    \left\langle M_\phi \right\rangle = -\left\langle \frac{\partial H}{\partial \phi} \right\rangle = H_\textbf{r} \left\langle S^y \right\rangle  \sin(\theta).
\end{equation}
The generalized force for a spin in a magnetic field is zero in the adiabatic limit, as the spin will simply align with the field. As such, the Berry curvature can be found as the leading order correction with eq. (\ref{eq:forcecurve}). We obtain that the Berry curvature can be calculated as
\begin{equation} \label{eq:17}
    F_{\theta \phi}  = F_{\theta 0}  = \frac{H_\textbf{r}}{v_\theta} \sin(\theta)\left\langle S^y \right\rangle.
\end{equation}
The Chern number is found by integrating the Berry curvature over the entire manifold. However, since the Hamiltonian is cylindrically symmetric, we need only integrate over $\theta$. The Chern number can then be found as
\begin{equation} \label{eq:chfthph}
    \ch = \frac{1}{2\pi} \int_0^\pi \int_0^{2\pi} F_{\theta\phi} d\phi d\theta =
    \int_0^\pi F_{\theta\phi} d\theta.
\end{equation}
In a physical experiment, $\theta$ would be ramped linearly for a time between $0$ and $t_\text{ramp}$, then $\left\langle S^y \right\rangle$ would be measured and the experiment would be performed again, ramping to a different time, so that the evolution can be tracked.

Eq. (\ref{eq:chfthph}) gives a prescription of the Chern number in terms of the Berry curvature. However, since the number is a description of the number of enclosed degeneracies in parameter space, we can also find these analytically. The path described previously is on the surface of a spherical manifold of radius $H_\textbf{r}$. Thus, in the simple case described here, the Weyl point lies inside the sphere for $H_0 < H_\textbf{r}$ and outside for $H_0 > H_\textbf{r}$. Comparing with eq. (\ref{eq:fipi}), we see that the Chern number is $\ch=2$ when $H_0 < H_\textbf{r}$. Simulating the ramping of $\theta$ and integrating $\left\langle F_{\theta\phi} \right\rangle$ over the resulting semicircle yields a numerical way to calculate $\ch$.

\begin{figure}
    \begin{minipage}[c]{0.9\columnwidth}
			\includegraphics[width=\columnwidth]{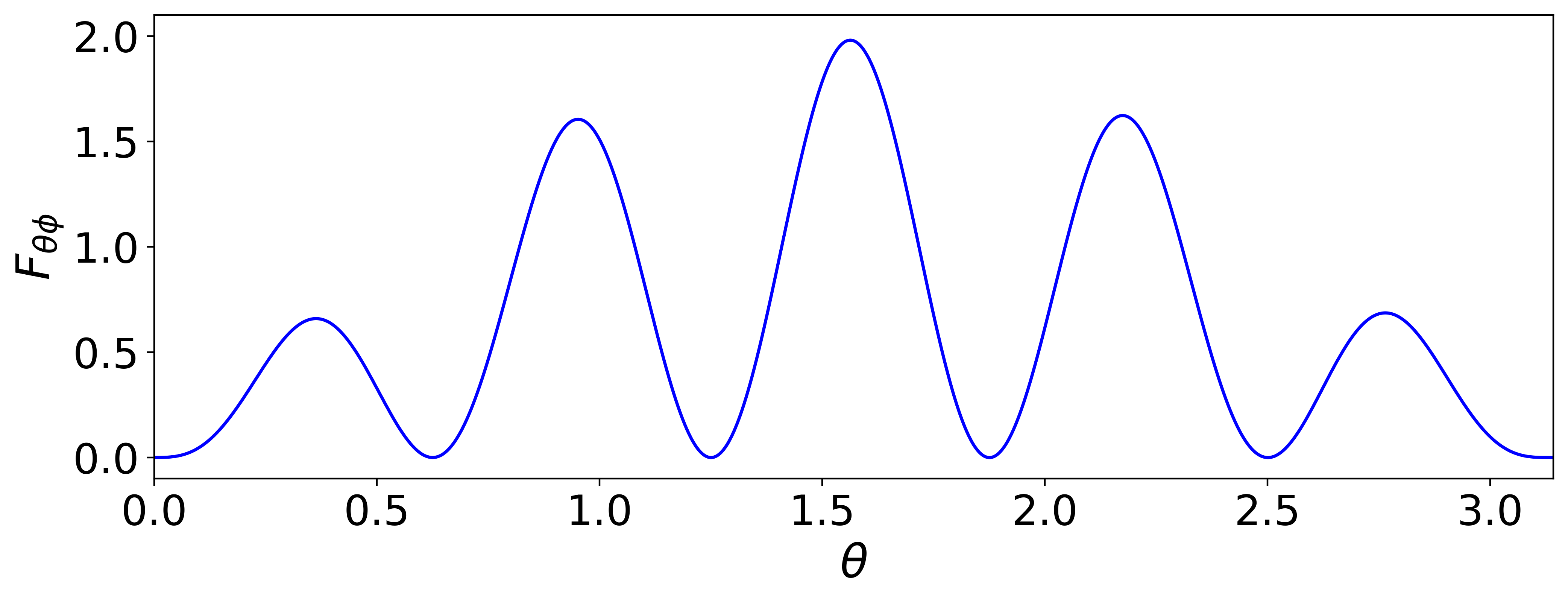}
		\end{minipage}
		\subfloat[]{\label{fig:Fthph}}
		\\[1ex]
		\begin{minipage}[c]{0.9\columnwidth}
			\includegraphics[width=\columnwidth]{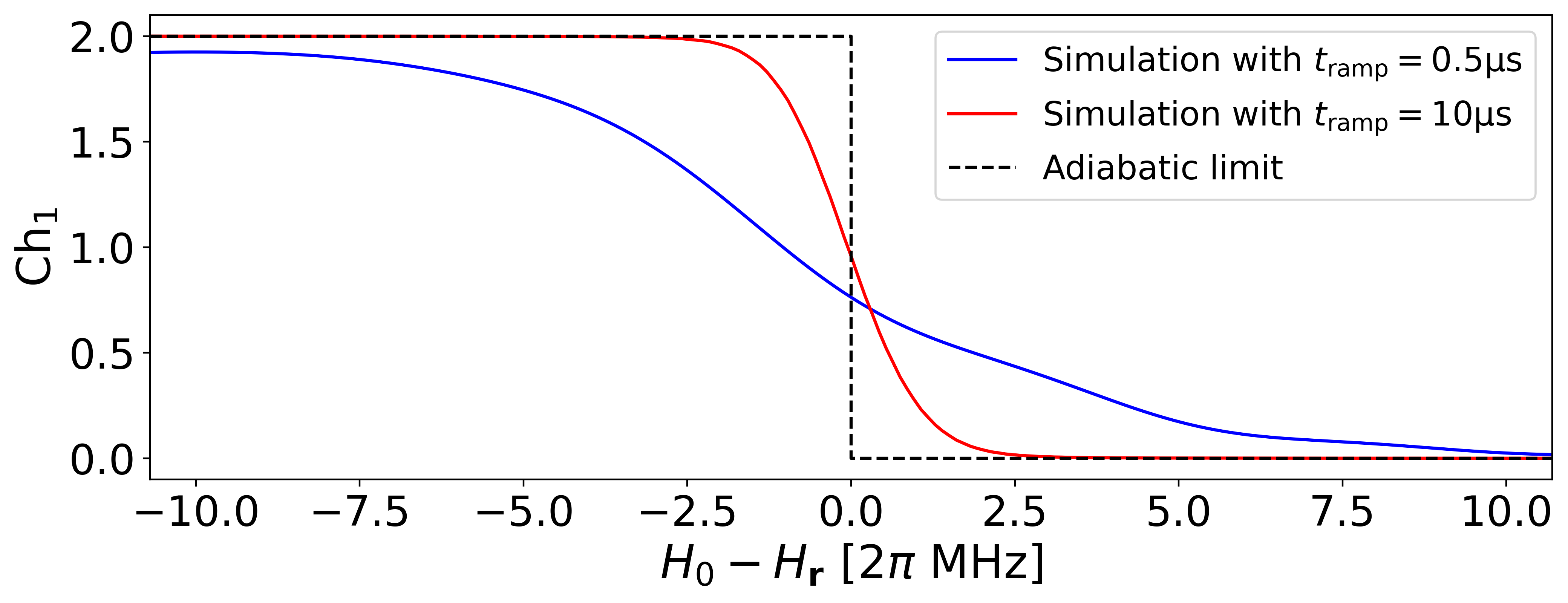}
		\end{minipage}
		\subfloat[]{\label{fig:transition}}
		\\[1ex]
		\begin{minipage}[c]{0.45\columnwidth}
			\includegraphics[width=\columnwidth]{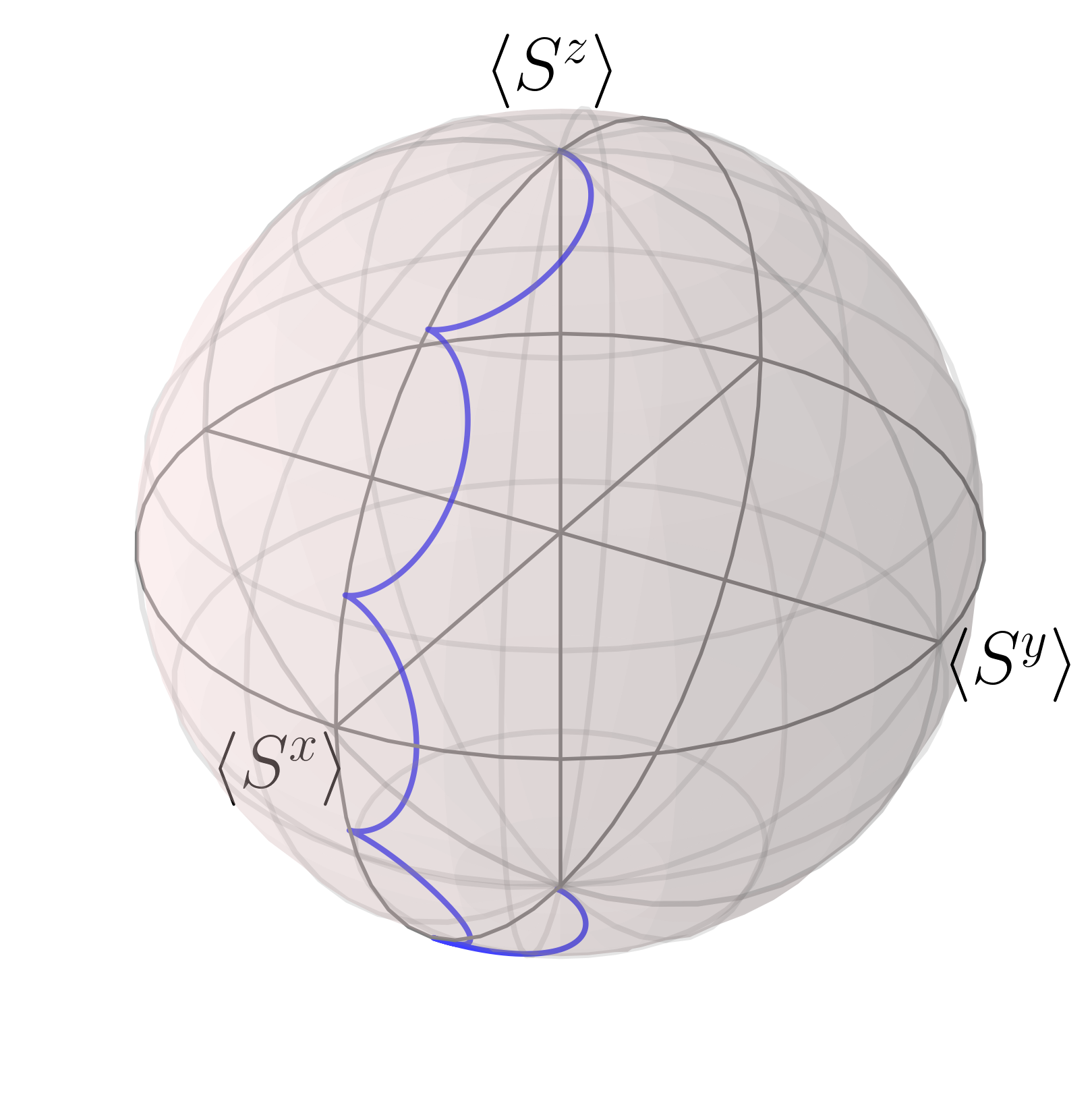}
		\end{minipage}
		\subfloat[]{\label{fig:traject}}
    \caption{\textbf{(a)} Plot of the Berry curvature $F_{\theta\phi}(\theta)$ for the case $H_0=0$ with $t_\text{ramp}=\SI{0.5}{\micro\second}$ and $ H_{\bm{r}} = 10 \cdot 2\pi\ \si{\mega\hertz}$. The integral of $F_{\theta\phi}$ is used to find $\ch$. 
    \textbf{(b)} $\ch$ for various values of $H_0$, with $t_\text{ramp}=\SI{0.5}{\micro\second}$ and with $t_\text{ramp}=\SI{10}{\micro\second}$. A topological transition occurs when $H_0=H_\textbf{r} = 10 \cdot 2\pi\ \si{\mega\hertz}$, where the Weyl point is at the boundary of the ground state manifold. The probing of the topological transition is made clearer in the second case due to the slow ramping.
    \textbf{(c)} The trajectory of the state on the surface of a sphere spanned by $\left\langle S^x \right\rangle$, $\left\langle S^y \right\rangle$ and $\left\langle S^z \right\rangle$ with $t_\text{ramp}=\SI{0.5}{\micro\second}$. Starting from $\left\langle S^z \right\rangle=1$ at $t=0$ to $\left\langle S^z \right\rangle=-1$ at $t=t_\text{ramp}$. The deviation from the meridian is used to extract the Berry curvature and the Chern number.}
    \label{fig:Berry}
\end{figure}

For the sake of example, the ramping of $\theta$ is  performed with  $t_\text{ramp}=\SI{0.5}{\micro\second}$. For $H_0=0$ this results in the $F_{\theta\phi}$ plotted in fig. \ref{fig:Fthph}. 
In this case the integral gives $\ch\approx 2$, since the Weyl point is enclosed. The topological nature of the number can be investigated by simulating for different values of $H_0$. The result of these simulations is shown in blue in fig. \ref{fig:transition}. While it is clear that $\ch$ changes from $2$ to $0$, it does so continuously. This is a result of the measurement technique, as higher order response from eq. (\ref{eq:forcecurve}) makes eq. (\ref{eq:17}) invalid. If we instead choose to simulate with $t_\text{ramp}=\SI{10}{\micro\second}$, we get a noticeably sharper transition plotted in red in fig. \ref{fig:transition}.
As expected, a transition occurs at the point $H_0 = H_\textbf{r}$, with $\text{Ch}_1=2$ when $H_0 < H_\textbf{r}$ and $\text{Ch}_1=0$ when $H_0 > H_\textbf{r}$.

The number of oscillations in $F_{\theta\phi}$ is given by $H_\textbf{r}/ v_\theta$. As a result of this, choosing a slower ramping speed will result in a higher number of oscillations and therefore a higher resolution in time is needed to accurately determine $\ch$. Since the integral of $F_{\theta\phi}$ over the sphere is independent of $t_\text{ramp}$, eq. (\ref{eq:17}) also means that $\expval{S^y}$ becomes less pronounced, requiring greater precision in its measurement.
On the other hand, higher levels in the superconducting circuit spectrum are more unstable \cite{peterer2015coherence}. As a result of this, a fast ramp may then mitigate some of the decay noise that might otherwise occur in slow transition protocols.
Due to these considerations, we set $H_\textbf{r}=\SI{10}{\mega\hertz}$ and  $t_\text{ramp}=\SI{0.5}{\micro\second}$ for the remainder of the paper. This is also consistent with the energy scale of the Hamiltonian of superconducting circuit relevance and in the same order as similar experiments \cite{schroer,roushan,Tan18}.

In the case of a qubit system, the response can be visualised as a trajectory on a Bloch sphere. Since we are dealing with a qutrit, more parameters are necessary to completely describe the state of the system. We can however, still visualise the relevant parts of the trajectory, described by $\left\langle S^x \right\rangle$, $\left\langle S^y \right\rangle$ and $\left\langle S^z \right\rangle$. This can be considered as a cross section of the entire space. This trajectory with $t_\text{ramp}=\SI{0.5}{\micro\second}$ for the case $H_0=0$ is plotted in fig. \ref{fig:traject}.

\section{Coupled three-level system}
\subsection{Introduction to the coupled three-level system} \label{sec:coupthree}
Having discussed the single three-level system, we now continue to the coupled case. The modification we perform is to add an additional qutrit system coupled to the first with a coupling term $g (S_1^x S_2^x+S_1^y S_2^y)$. This is the generalization to higher spins of the coupled spin 1/2 system as for 
instance discussed in \cite{roushan}. Here we will explore how the extra levels and the interactions can 
influence the phase diagram. 

We consider Hamiltonians of the form
\begin{equation} \label{eq:coupham}
    H= -\left(H_{0} S^z_1 + \textbf{H}_1\cdot\textbf{S}_1 + \textbf{H}_2\cdot\textbf{S}_2 - g (S_1^x S_2^x+S_1^y S_2^y)\right),
\end{equation}
where we will set $\textbf{H}_1=\textbf{H}_2=\textbf{H}_\textbf{r}$, again with $$\textbf{H}_\textbf{r}=H_\textbf{r}\left( \sin(\theta)\cos(\phi), \sin(\theta)\sin(\phi), \cos(\theta) \right),$$ 
with $H_\textbf{r}=10\cdot 2\pi$ MHz. As before, the system is initialized in its instantaneous ground state for a fixed value of $H_0$ and $g$. We ramp $\theta(t)$, this time for both qutrits simultaneously, with both $H_{0}$ and $g$ kept constant. Since $g$ is kept constant for each simulation, by taking the $\phi$-derivative of (\ref{eq:coupham}), we see that the Berry curvature is obtained as the integral over two separate spherical manifolds. Likewise, the system has $U(1)$ invariance along the $z$ axis, meaning that the system for $H(\theta, 0)$ can be mapped to $H(\theta, \phi)$ as
\begin{equation}
    H(\theta, \phi) = e^{i\phi\left(S^z_1+S^z_2\right)} H(\theta, 0) e^{-i\phi\left(S^z_1+S^z_2\right)}.
\end{equation}
With this in mind, the Chern number is simply found by integrating the Berry curvature for both qutrits
\begin{equation} \label{eq:2cherns}
    \ch = \int_0^\pi \frac{H_\textbf{r}}{v_\theta}  \sin(\theta) \left( \left\langle S_1^y \right\rangle + \left\langle S_2^y \right\rangle \right) d\theta
\end{equation}

In the limiting case of $g=0$, we expect this to be equivalent to two isolated qutrits, one of which has a transition where $\ch$ changes by 2 when $H_0=H_\textbf{r}$, while the other is unaffected by the change in $H_0$, i.e., we expect $\ch=4$ when $H_0<H_\textbf{r}$ and $\ch=2$ when $H_0>H_\textbf{r}$.
When $g\neq 0$, things become more complicated. We expect multiple phases of the system to be present, with different values of $\ch$ depending on the strength of $H_0$ and $g$.
The theoretical phase diagram can again be found by considering the position of degeneracies in the parameter space. We expect the total $\ch$ to be no larger than the sum of the individual $\ch$ for each isolated qutrit, since the coupling and offset is kept constant for each ramp. That is, we expect $\ch\leq 4$, no matter the value of $H_0$ and $g$.
\subsection{Analytical solution}
We first reorient our Hamiltonian to lie in the $\hat{\textbf{z}}$-direction, so that $\theta=0$ or $\theta=\pi$. We can analytically locate the Weyl points of the system by considering this situation. For $\theta=0$,  the Hamiltonian, in the basis $\{\uparrow,0,\downarrow\} \otimes \{\uparrow,0,\downarrow\}$, reduces to
\begin{widetext}
\begin{equation}
H = 
    \begin{pmatrix}
-H_{0}-2 H_\textbf{r} & 0 & 0 & 0 & 0 & 0 & 0 & 0 & 0 \\
0 & -H_{0}-H_\textbf{r} & 0 & g & 0 & 0 & 0 & 0 & 0 \\
0 & 0 & -H_{0} & 0 & g & 0 & 0 & 0 & 0 \\
0 & g & 0 & -H_\textbf{r} & 0 & 0 & 0 & 0 & 0 \\
0 & 0 & g & 0 & 0 & 0 & g & 0 & 0 \\
0 & 0 & 0 & 0 & 0 & H_\textbf{r} & 0 & g & 0 \\
0 & 0 & 0 & 0 & g & 0 & H_{0} & 0 & 0 \\
0 & 0 & 0 & 0 & 0 & g & 0 & H_{0}+H_\textbf{r} & 0 \\
0 & 0 & 0 & 0 & 0 & 0 & 0 & 0 & H_{0}+2 H_\textbf{r}
    \end{pmatrix}.
\end{equation}
\end{widetext}
We will be localizing Weyl points in the parameter space, as we know from eq. \eqref{eq:CurvNab} that this is the location in parameter space that causes a non-trivial Berry curvature. The only relevant coordinate of the Weyl points, the $z$-coordinate, will be denoted $H_z$.

We block-diagonalize the Hamiltonian in the eigenbasis of $J_z$. As such, we analyze the system in terms of each value of the total spin along $z$, $m_{tot}$.

For $m_{tot}=\pm 2$, we have energies $E_{\uparrow \uparrow}=-H_0-2 H_\textbf{r}$ and $E_{\downarrow \downarrow}=H_0+2 H_\textbf{r}$.
For $m_{tot}=\pm 1$ we have the matrices
\begin{equation}
    H_{0 \uparrow / \uparrow 0} = 
    \begin{pmatrix}
    -H_0-H_\textbf{r} & g\\
    g & -H_\textbf{r}
    \end{pmatrix}, \ \ H_{\downarrow 0 / 0 \downarrow} = 
    \begin{pmatrix}
    H_\textbf{r} & g\\
    g & H_0+H_\textbf{r}
    \end{pmatrix},
\end{equation}
with energies $\frac{-H_{0}}{2}-H_\textbf{r}\pm\frac{\sqrt{H_{0}^{2}+4 g^{2}}}{2}$ and $\frac{H_{0}}{2}+H_\textbf{r}\pm\frac{\sqrt{H_{0}^{2}+4 g^{2}}}{2}$, respectively.
Finally the $m_{tot}=0$ sector looks like
\begin{equation}
    H_{\uparrow \downarrow / 00 / \downarrow \uparrow}
    =
    \begin{pmatrix}
    -H_0 & g & 0 \\
    g & 0 & g \\
    0 & g & H_0
    \end{pmatrix}
\end{equation}
With energies $0$ and $\pm \sqrt{H_{0}^{2}+2 g^{2}}$.
Since we have $\nabla_\mathbf{R}H = \textbf{S}=\frac{1}{2}\left(S^+ + S^-\right)\hat{\textbf{x}} + \frac{1}{2i}\left(S^+ - S^-\right)\hat{\textbf{y}} + S_z \hat{\textbf{z}}$ and since it is an orthogonal basis, eq. (\ref{eq:CurvNab}) gives that, only states that differ in $m_{tot}$ by $1$, will be able to contribute to the Berry curvature. That is, degeneracy from states with a difference in spin other than 1, are irrelevant.

Depending on the values of $H_0$ and $g$, there will be 3 different ground states, these can be found to be
\begin{equation}
\ket{g} = 
    \begin{cases}
        \mid \uparrow\uparrow\rangle\\
        a_1\mid\uparrow \! 0\rangle - a_2 |0\!\uparrow\rangle\\
        b_1|00\rangle - b_2\mid\uparrow \downarrow\rangle - b_3\mid\downarrow\uparrow\rangle
    \end{cases}
\end{equation}
where $a_i$ and $b_i$ are positive constants.
In the case $H_0=0$, we get in particular $a_1=a_2=b_2=b_3=1/\sqrt{2}$ and $b_1=1/2$. 
As such, the ground state Berry curvature will be different in each case.

We first consider $\textbf{F}_{\uparrow\uparrow}$. Here the curvature will be stemming from degeneracies between $m_{tot}=+2$ and $m_{tot}=+1$, the two resulting ground state degeneracies have $z$-components
\begin{equation}
    H_z \in \left\{-\frac{H_{0}}{2}\pm\frac{1}{2}\sqrt{H_{0}^{2}+4 g^{2}} \right\}
\end{equation}

\begin{figure}
    \centering
    \includegraphics[width=\columnwidth]{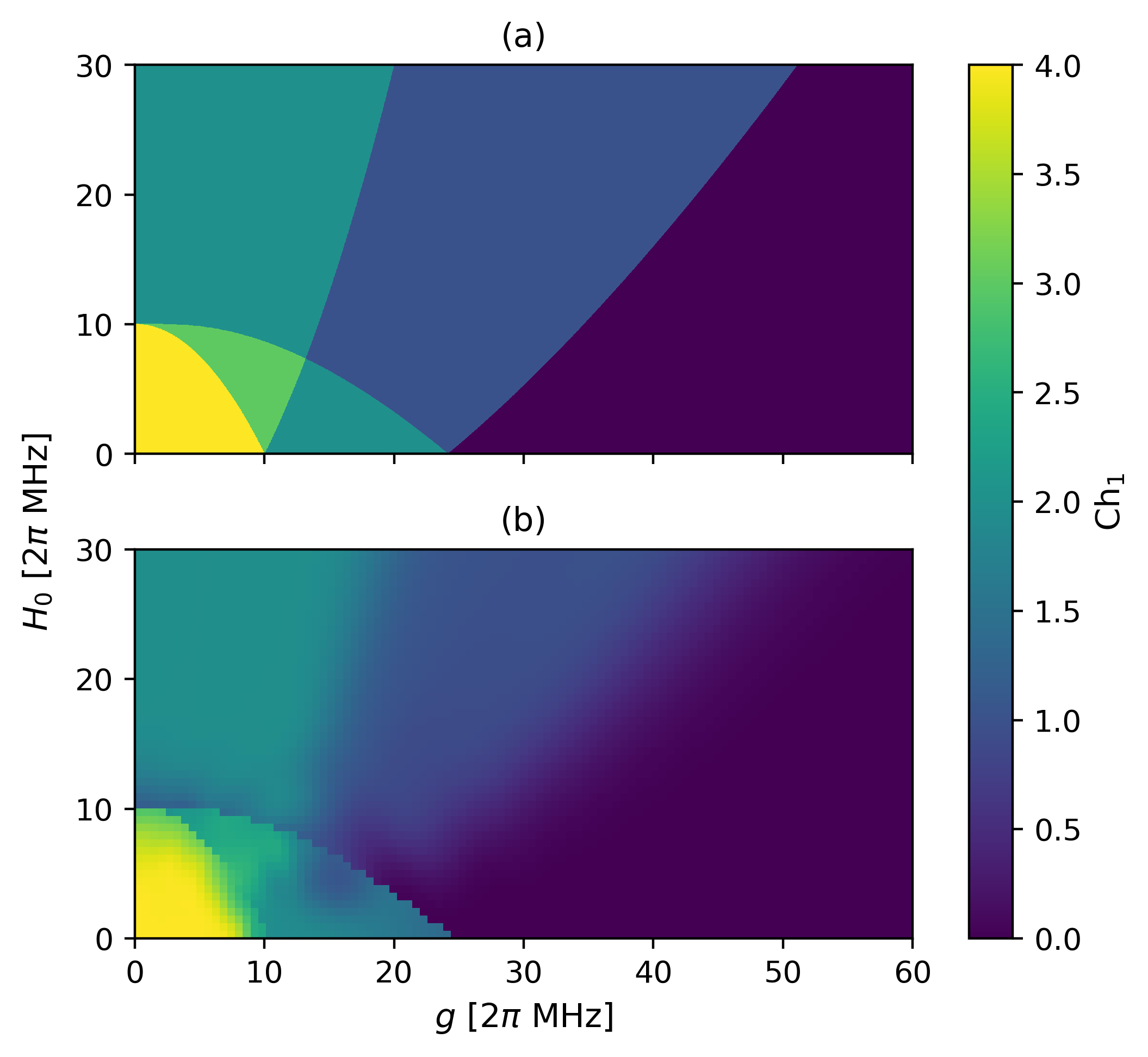}
    \caption{\textbf{(a)} The theoretical phase diagram. 6 different phases are present, with different values of $\ch$. This was constructed by considering how many degeneracies are enclosed within the ground state manifold with $ H_{\bm{r}} = 10 \cdot 2\pi\ \si{\mega\hertz} $.
    \textbf{(b)} Simulated phase diagram, showing the value of $\ch$, which was found by integrating $F_{\theta\phi}$, for different values of $H_0$ and $g$, with $H_{\bm{r}} = 10 \cdot 2\pi\ \si{\mega\hertz}$ and $t_\text{ramp} = \SI{0.5}{\micro\second}$.}
    \label{fig:H0gtheo}
\end{figure}

We now consider $\mathbf{F}_{\uparrow0/0\uparrow}$, the curvature from the states with $m_{tot}=+1$. Here, the relevant degeneracies are between the $m_{tot}=+1$ sector and both the $m_{tot}=+2$ and $m_{tot}=0$ sector. The new ground state degeneracies have $z$-components
\begin{equation}
    H_z \in \left\{ -\frac{H_{0}}{2} \pm \frac{1}{2}\sqrt{H_{0}^{2}+4 g^{2}} \mp \sqrt{H_{0}^{2}+2 g^{2}} \right\}
\end{equation}
All remaining values of $H_z$ will, up to a sign, be identical to the ones already found. This can also be seen from our previous argument that $\ch\leq4$, with four values of $H_z$ having been found.
In total, the 4 relevant locations are
\begin{equation}
    \begin{aligned}
    H_z \in &\left\{-\frac{H_0}{2}\pm\frac{1}{2} \sqrt{H_{0}^{2}+4 g^{2}},\right. \\
    & \ \left.-\frac{H_{0}}{2}\pm\frac{1}{2} \sqrt{H_{0}^{2}+4 g^{2}} \mp \sqrt{H_{0}^{2}+2 g^{2}}\right\}
    \end{aligned}
\end{equation}
These are the $z$-components of the Weyl points.
From this, we can construct a theoretical phase diagram. This is done in fig. \ref{fig:H0gtheo}a.
As is evident in the figure, six different phases are present, depending on how many $H_z<|H_\textbf{r}|$, i.e. how many of the degeneracies are enclosed within the sphere.

\subsection{Simulated solution}
As with the isolated qutrit, this system has been simulated for different values of $H_0$ and $g$. $\ch$ is again found by integrating $\left\langle S^y \right\rangle$ over the sphere, this time for both qutrits as in eq. (\ref{eq:2cherns}).
This phase diagram is plotted in fig. \ref{fig:H0gtheo}b.
The phase diagram is very interesting for a number of reasons. First of all, it is interesting, that while the uncoupled system only has a single transition where $\ch$ changes by 2, this system has a lot of possible transitions where $\ch$ can change by 1. For a fixed value of $g$ between 0 and $H_\text{r}$, changing $H_0$ from 0 to $H_0>H_\text{r}$ will still cause $\ch$ to change by $2$, but in two steps. In section \ref{sec:simple}, the change in $\ch$ by $2$ could be interpreted as moving the single monopole with charge $2$ from inside to outside the ground state manifold. This interpretation is no longer valid, given how it changes in two steps. Instead one must interpret it as two distinct monopoles with charge $1$.
Likewise, for $H_0=0$, changing $g$ allows for two different transitions, while for $H_0$ between $0$ and $H_\text{r}$, four transitions occur.

The fact that the phases seem to "split" as $g$ and $H_0$ increases, is interesting in itself. 
The fact that there exists phases that are not driven purely by $H_0$ and $g$, but for specific values of these in combination, is in itself a rather remarkable result. This result is not possible in the spin-$1/2$ analogue of the system, given that each particle only contribute with $\ch=1$ each, only three phases are present \cite{roushan}. This system, therefore, allows for completely new types of phases to occur that would not be possible for lower spin.

\section{Implementation of coupled qutrit system}
\label{sec:three-level}
\begin{figure}
    \centering
    \includegraphics[width=0.8\columnwidth]{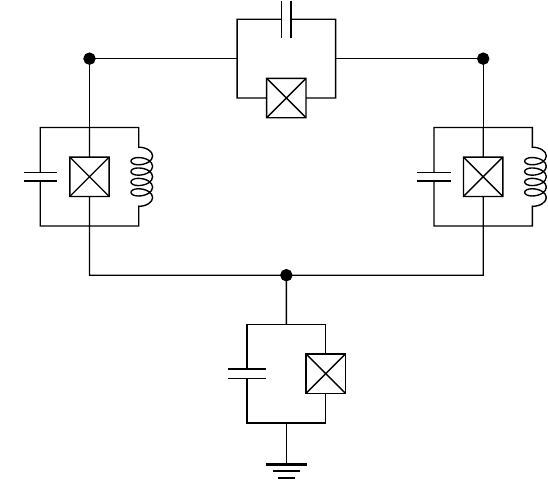}
    \caption{A lumped-element circuit drawing of the proposed superconducting circuit implementing a coupled two-qutrit system with an effective Heisenberg $XXZ$ interaction. Standard circuit notation is used, with the crossed boxes representing Josephson junction.}
    \label{fig:circuit}
\end{figure}

As an example of how a system of two coupled spin 1 systems, i.e. qutrits, could be implemented in a physical device, we look at the superconducting circuit proposed in \cite{rasmussen2021superconducting}, section XI, D.
A lumped circuit element drawing of the system, can be seen in fig. \ref{fig:circuit}. The system consists of three superconducting islands indicated by nodes in the figure. The islands are connected using Josephson junctions, capacitors and inductors, while the bottom island is connected to the ground as a Transmon \cite{PhysRevA.76.042319_Transmon}. The two effective coupled degrees of freedom can be seen as the dipoles fluctuating between the upper two islands, and between the lower island and equally to each upper island, respectively. The third degree of freedom is a combined fluctuation with respect to the ground and can thus be ignored. The ratio of the inductances and capacitances are chosen such that the two dipoles are in the transmon regime and can thus be truncated to the lowest three levels. The Hamiltonian from this circuit has the following form
\begin{alignat}{1} \label{eq:SpinHam}
    H & = g\left(S^x_1S^x_2+S^y_1S^y_2\right)+J_Z\left(S^z_1S^z_2\right)+ \textbf{H}_1 \cdot \textbf{S}_1 + \textbf{H}_2 \cdot \textbf{S}_2  \nonumber\\ &  +
    H_0 S_1^z + \frac{J_{02}}{4}\left( \left(S^+_1 S^-_2\right)^2 + \left(S^-_1 S^+_2\right)^2 \right).
\end{alignat}
A deeper explanation of how the Hamiltonian arises is beyond the scope of this paper, however in appendix \ref{sec:app-sc} we describe briefly how the Hamiltonian of the system can be found, and how to modify it so that it is written in terms of spin-1 matrices. Further details can be found in Refs. \cite{rasmussen2021superconducting} and \cite{Baekkegaard}, including a discussion of the parameters involved. Intuitively, the two effective degrees of freedom can be seen as the electrical field modes oscillating between the left and the right upper circuit nodes, and oscillating between the two upper circuit nodes (black dots in \ref{fig:circuit}) and the bottom circuit node, respectively.

We note here that the $J_{02}$ term is a non-trivial term that appears here because of 
nature of the superconducting circuit setup when it is driven in order to make the parameters configurable. We therefore need to ensure that this will not interfere with the topological properties and the phase diagram. Luckily, this turns out not to be the case as we'll will discuss in the next section.

\section{Exploring the larger parameter space}
\label{sec:fourspin}
The new system is considerably more complicated than the one for a simple coupling, described in section \ref{sec:coupthree}. In the regime where $J_z \to 0$ and $J_{02} \to 0$, it reduces to the previously described system. It is quickly noted that this system also retains the $U(1)$ invariance, as can be seen by writing out 
\begin{equation}
    H(\theta, \phi) = e^{i\phi\left(S^z_1+S^z_2\right)} H(\theta, 0) e^{-i\phi\left(S^z_1+S^z_2\right)}.
\end{equation}
in matrix form.

In the case of the simple coupling in section \ref{sec:coupthree}, the interaction was controlled by $S^x$, $S^y$, $S^z$ for both qutrits and $\left( S_1^x S_2^x + S_1^y S_2^y \right)$, giving a parameter space of dimension seven. Adding control of $S_1^z S_2^z$ and $\left(\left(S_1^+ S_2^-\right)^2+\left(S_1^- S_2^+\right)^2\right)$ expands this to nine. Changing both the previous free variables, as well as the two new ones, would require 4-dimensions phase diagrams. For convenience, we will only be plotting 2-dimensional slices of this. 

As before, we will be considering spherical manifolds with constant $|\textbf{H}_1|=|\textbf{H}_2|=H_\textbf{r}=10\cdot 2\pi$ MHz. 

\begin{figure}
    \centering
    \includegraphics[width=\columnwidth]{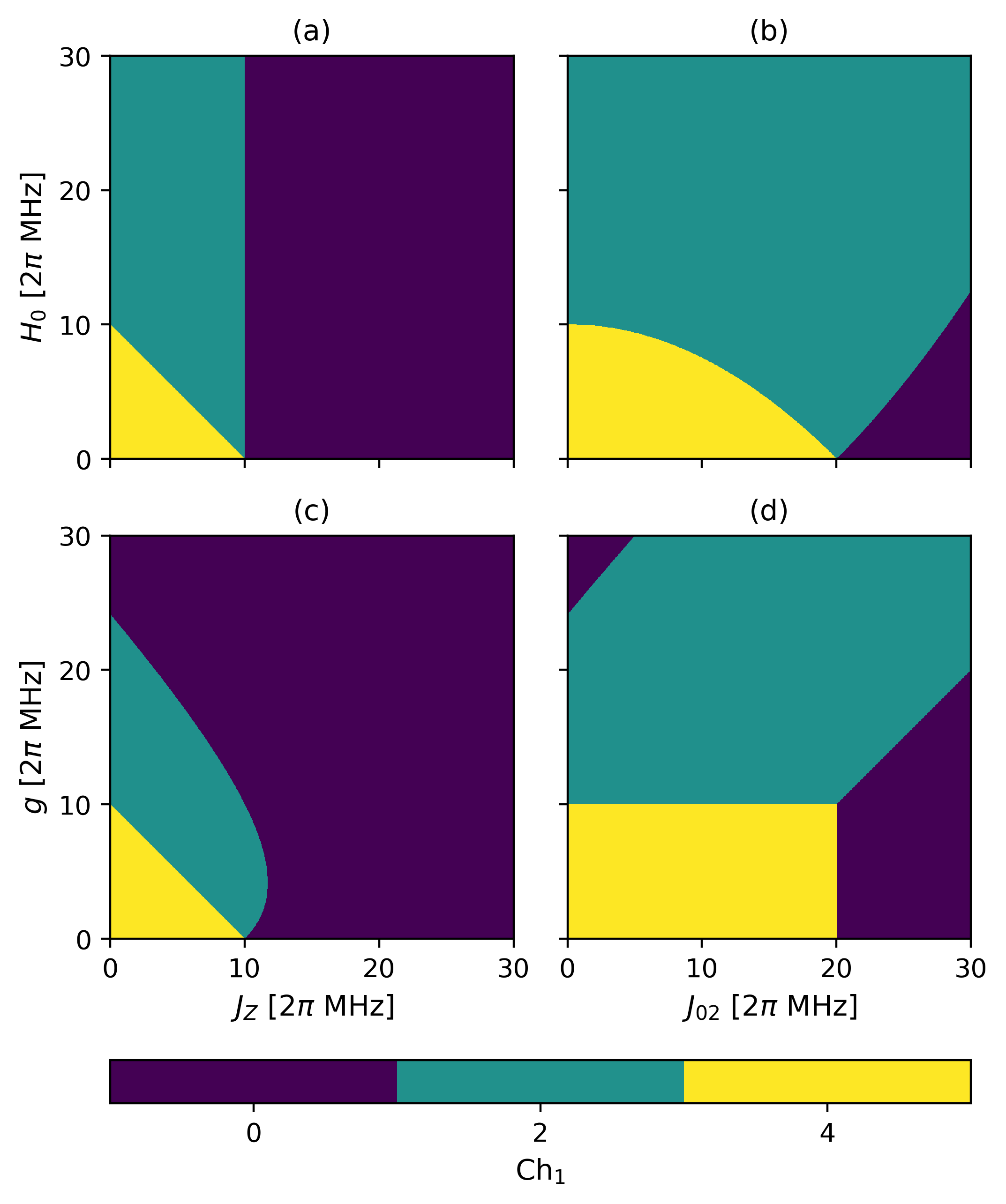}
    \caption{Phase diagram for $\ch$ with $H_\textbf{r} = 10 \cdot 2\pi\ \si{\mega\hertz}$ for various types of coupling with all others absent. \textbf{(a)} $H_0$ and $J_Z$. \textbf{(b)} $H_0$ and $J_{02}$. \textbf{(c)} $g$ and $J_{Z}$. \textbf{(d)} $g$ and $J_{02}$.}
    \label{fig:allcoups}
\end{figure}

In the simple case of only two parameters, we can easily construct theoretical phase diagrams in a similar fashion to that of section \ref{sec:coupthree}. We will do so with the new coupling parameters. We will not provide in-depth derivation of the theoretical phase diagrams, since the procedure is identical to the one previously applied. That is, first reorient the axis to point in the $\hat{\textbf{z}}$-direction, find ground state wave functions for each region, then locate the Weyl points. We have chosen to do this with a selection of the parameters, collected in fig. \ref{fig:allcoups}. 

Simulated diagrams can also be constructed by ramping $\theta(t)$ linearly and integrating $\left\langle S^y \right\rangle$ for each qutrit. This will not be presented. However, we note that all the theoretical values agree with simulations. The remainder of this section will be spent discussing the phase diagrams.

We first consider the effect of the $J_Z$ coupling. The resulting diagram is plotted in fig. \ref{fig:allcoups}a. As is evident in the figure, three phases are present, with transitions occurring at $H_0+J_Z=H_\textbf{r}$ and $J_Z=H_\textbf{r}$. This phase diagram is quite different than the one for $g$-coupling, as the transitions are much simpler in this regime. The fact that $H_0$ and $J_Z$ can both contribute equally to the transition where $\ch$ changes from $4$ to $2$, but only the coupling induces the transition where $\ch\to0$, is interesting. Alone however, the diagram is not so illuminating.

Next, we focus on the effect of varying the parameters $J_{02}$ and $H_0$. This is presented in fig. \ref{fig:allcoups}b.
It is worth noting that the phase diagram produced by this coupling is identical to the first phase of the one controlled by $g$-coupling, except with a scaling factor on the coupling parameter. To be exact, the $z$-component of the Weyl point is
$$
    H_z = -\frac{H_0}{2}\pm\frac{1}{2}\sqrt{H_0^2+J_{02}^2}
$$
compared to
$$
    H_z = -\frac{H_0}{2}\pm\frac{1}{2}\sqrt{H_0^2+4g^2}
$$
for $g$-controlled coupling.

The dynamics that are most common in spin chains, are those governed by the coupling of the types $(S_1^x S_2^x+S_1^y S_2^y)$ and $(S_1^z S_2^z)$, controlled by $g$ and $J_Z$.
Phase diagrams for the two couplings controlled by $J_Z$ and $g$ is plotted in fig. \ref{fig:allcoups}c.
As in the case values of $J_Z$ and $H_0$, the first transition is controlled equally by the two parameters, while the second transition is of a more complex nature. 

Finally, we focus on the effect of $J_{02}$ and $g$ as in fig. \ref{fig:allcoups}d.
It can be seen that when $g<H_\textbf{r}$ and $J_{02}<2H_\textbf{r}$, neither coupling is strong enough to change the topology, but that topological transitions occur when $g=H_\textbf{r}$ or $J_{02}=2H_\textbf{r}$.
The most interesting thing is perhaps that while a strong increase in either $g$ or $J_{02}$ will result in $\ch\to 0$, we have $\ch=2$ when $J_{02}\approx g > H_\text{r}$, no matter their magnitude.

The phase diagrams presented here are very rich. This indicates that a relatively simple generalization of the case of two coupled spin 1/2 systems, as performed in \cite{roushan}, can give a much more varied set of topological phases as one tunes the parameters. We note that this comes at the expense of adding a single extra level to the superconducting circuits. Exploring these new phase diagrams in greater detail than what has been done here, by for example interpreting the phases and examining properties such as critical exponents of the topological transitions should be a venue for future work. 

Furthermore, the lifetimes of higher states beyond first two are typically only about a factor of two shorter. Compared to \cite{schroer} where a ramping for a two-level system for 1 microsecond was performed, we expect that the three-level system can be explored in the same manner, with ramping times of the same order as explored in the simulations, i.e. at the level of around 0.5 microseconds. 
This may yield further insight into coupled higher-spin systems as a means of simulating topological phases in quantum systems.

\section{Conclusion and outlook}
\label{chap:conclusion}
In this paper, systems of interacting spins beyond the simple spin 1/2 system have been considered. 
Starting from a single spin-$1$ system, we introduced the theoretical groundwork in terms of spin-$1$ operators, Berry curvature, and the first Chern number in order to address the topological transitions that are possible for systems with higher spins. Furthermore, we also connected these considerations to previous measurement protocols used to probe different topological phases, and how to induce 
transitions between different phases by changing system parameters. 
We then proceeded to add a second spin-$1$ system and studied the coupling of two such qutrits. This turns out to produce a considerably richer phase diagram compared to the case of coupled spin 1/2 systems. Finally, a system based on the platform of superconducting circuits was introduced that can be used to realize a set of coupled spin 1 systems, and suitably modified with drive lines in order to obtain control of different system parameters. The system was simulated for a large set of parameters and the ensuing dynamics was characterized. This allowed us to conclude that a lot of opportunities may lie in the experimental probing of higher spins for its topological properties. 

Our analysis here did not take into account some issues that may arise in realizations with superconducting circuits. The issue of how noise may effect the phases, and how to tune the driving or optimize the protocols in order to mitigate noise effects was left for future consideration. 

Topological properties are of increasing interest in several fields of physics, since they allow for unique physical phenomena that typically goes beyond our basic intuitions, and the realization of topological phases in quantum simulators and in materials have become a very intensely pursued topic. 
In order to derive topological features of different physical systems, it is common to define mappings onto systems where topological features are easier to identify. Quite often, this will be to forms that are closely related to spins in magnetic fields. There have been some works along these lines, see for instance \cite{roushan}, \cite{Zhang2017} and \cite{Tan18}. We therefore speculate that the three-level systems coupled with $J_{02}$- and $J_Z$-terms, can be mapped to a physical system, in which the topological phases of section \ref{sec:fourspin} become manifest. 
Furthermore, as was noted in \cite{Baekkegaard}, it is quite simple to imagine how to scale 
the present system of two qutrits to a larger chain of coupled qutrits using the superconducting circuit designs that lead to the Hamiltonian considered here. As is briefly mentioned at the end of \cite{Tan18}, this could be used to implement the Haldane phase of an interacting spin-1 chain \cite{PhysRevLett.50.1153}. However, we note that some of the other platforms mentioned in the introduction that are currently in use for topological phase quantum simulations may also serve this purpose. Hence, the principle theoretical investigation started here with coupling of two spin 1 systems could serve as a 'few-body' precursor to the realization of many-body phases with manifest topological properties.  

\appendix

\section{Superconducting circuits}
\label{sec:app-sc}
In section \ref{sec:three-level} we introduced a Hamiltonian for a physical realization of a coupled three-level system, stemming from the circuit in fig. \ref{fig:circuit}. This appendix briefly presents the full coupled Hamiltonian investigated in \cite{Baekkegaard} and later in \cite{rasmussen2021superconducting} and describes how to modify it to arrive at eq. (\ref{eq:SpinHam}).

The circuit Hamiltonian can be derived by following the standard quantization procedure described in \cite{devoret-introduction, rasmussen2021superconducting}, see \cite{Baekkegaard, rasmussen2021superconducting}. As described in the main text, the three nodes in the circuit couple in such a way that a change of basis can be performed to a coordinate system where a center-of-mass like degree of freedom can then be ignored. The result is the Hamiltonian of effectively two coupled qutrits, denoted by subscripts $i=1$ and $i=2$: \begin{widetext}
\begin{equation}
\begin{aligned}
H_{\text {full }} & = \Delta_{1,1}|1\rangle \langle   1 |_{1} + (\Delta_{1,1} + \Delta_{1,2} ) | 2 \rangle \langle   2 |_{1}  \\
& +\Delta_{2,1}|1\rangle \langle   1 |_{2}+ (\Delta_{2,1}+\Delta_{2,2} ) | 2 \rangle \langle   2 |_{2} \\
&+J_{01,01} (|0\rangle \langle   1 |_{1} | 1 \rangle \langle   0 |_{2}+| 1 \rangle \langle   0 |_{1} | 0 \rangle \langle   1 |_{2} )+J_{01,12} (|0\rangle \langle   1 |_{1} | 2 \rangle \langle   1 |_{2}+| 1 \rangle \langle   0 |_{1} | 1 \rangle \langle   2 |_{2})      \\
&+J_{12,01} (|1\rangle \langle   2 |_{1} | 1 \rangle \langle   0 |_{2}+| 2 \rangle \langle   1 |_{1} | 0 \rangle \langle   1 |_{2} )+J_{12,12} (|1\rangle \langle   2 |_{1} | 2 \rangle \langle   1 |_{2}+| 2 \rangle \langle   1 |_{1} | 1 \rangle \langle   2 |_{2})      \\
&+J_{02} (|0\rangle \langle   2 |_{1} | 2 \rangle \langle   0 |_{2}+| 2 \rangle \langle   0 |_{1} | 0 \rangle \langle   2 |_{2} )  \\
&+J_{Z Z} (D_{1,1}|1\rangle \langle   1 |_{1}+D_{1,2} | 2 \rangle \langle   2 |_{1} ) (  D_{2,1}|1\rangle \langle 1 |_{2}+D_{2,2}|2\rangle \langle   2 |_{2} ),    \\
\end{aligned}
\end{equation}
\end{widetext}
using an outer product operator notation with the qutrits states denoted by 0, 1, and 2, respectively. The subscripts for $\Delta$, $D$ and $J$ are merely to distinguish them and to label which states they pertain to.
We note here that we have performed the rotating wave approximation in order to arrive at the Hamiltonian \cite{rasmussen2021superconducting}.
The structure of the coupling is similar to $XXZ$-type Heisenberg couplings \cite{loft2018quantum,christensen2020coherent}, but with some important modifications that is discussed below. These arise primarily from the driving terms required for full control of the system levels and the subsequent transformation to the rotating frame and diagonalization herein.

We will here make the simplifying assumption that all parameters can be externally controlled independently of each other. This would require some additional drive lines for multi-mode driving or other techniques added onto the original model. 
The procedure for rewriting has been done by writing out the terms in matrix form with the Kronecker product, and comparing with how the spin matrices are given in the same basis. Since the matrices are rather large, we will not write them out explicitly.
By setting $J_{01,01}=J_{01,12}=J_{12,01}=J_{12,12}=g$, lines 3 and 4 can be written 
\begin{equation*}
	g(S^x_1 S^x_2+S^y_1 S^y_2)
\end{equation*}
i.e. $XX+YY$-coupling that we considered in \ref{sec:coupthree}.
 
It is possible to remove a constant on the diagonal of the Hamiltonian in order to set the lowest state at
zero energy. 
Alternatively, we could throw away the middle $|1\rangle \langle 1|$-term and be left with 
$\Delta_{i,0}|0\rangle \langle 0 |_{i} + (\Delta_{i,0} + \Delta_{i,2} ) | 2 \rangle \langle   2 |_{i}$ for $i=1,2$ for both the terms acting directly on the qutrit and the coupling term. 
Setting $\Delta_{i,0} = -\Delta_{i,2}/2$, and similarly for the coupling term, gives rise to the terms
\begin{equation*}
\Delta_1 S^z_1 + \Delta_2 S^z_2 + J_{ZZ}(S^z_1 S^z_2).
\end{equation*}
Finally the $J_{02}$-term cannot immediately be written in terms of the $S^x$, $S^y$ or $S^z$ spin-1-operators. It can however be written in terms of the ladder operators as
\begin{equation*}
    J_{02}\left( \left(S^+_1S^-_2\right)^2 + \left(S^-_1S^+_2\right)^2 \right)/4
\end{equation*}
While not present in the presented circuit drawing (fig. \ref{fig:circuit}) for simplicity, adding terms giving control over $S^x$ and $S^y$ for both qutrits is a standard procedure. We refer to \cite{rasmussen2021superconducting} for details. Adding these terms, we have an implementation of the required spin-1 operators added to the Hamiltonian.

In the end we have the Hamiltonian
\begin{alignat}{1}
    H & = g\left(S^x_1S^x_2+S^y_1S^y_2\right)+J_Z\left(S^z_1S^z_2\right)+ \textbf{H}_1 \cdot \textbf{S}_1 + \textbf{H}_2 \cdot \textbf{S}_2  \nonumber\\ &  +
    H_0 S_1^z + \frac{J_{02}}{4}\left( \left(S^+_1 S^-_2\right)^2 + \left(S^-_1 S^+_2\right)^2 \right).
\end{alignat}
This is eq. (\ref{eq:SpinHam}) in the main text.

\bibliography{bibliography}

\end{document}